\documentclass[aip,jcp,reprint,floatfix]{revtex4-1}
\usepackage{verbatim}%
\usepackage{amsmath,amsfonts} %
\usepackage[acronym]{glossaries} %
\usepackage{graphicx} %
\usepackage{dcolumn} %
\usepackage{bm} %
\usepackage{ifpdf} %
\usepackage{color} %
\usepackage{nicefrac} %
\usepackage{natbib} %
\usepackage[byname]{smartref} %
\usepackage[breaklinks, %
colorlinks, %
linkcolor=blue, %
citecolor=blue, %
urlcolor=blue]{hyperref} %
\usepackage[table]{xcolor} %
\usepackage{braket} %

\newcommand{\eq}[1]{Eq.~(\ref{#1})} %
\newcommand{\eqs}[1]{Eqs.~(\ref{#1})} %
\newcommand{\bea}{\begin{eqnarray}}
\newcommand{\eea}{\end{eqnarray}}

\def\be{\begin{equation}}
\def\ee{\end{equation}}
\def\bea{\begin{eqnarray}}
\def\eea{\end{eqnarray}}
\def\bal#1\eal{\begin{align*}#1\end{align*}}
\def\ba#1\ea{\begin{align}#1\end{align}}

\newcommand{\fig}[1]{Fig.~\ref{#1}}
\newcommand{\figs}[1]{Figs.~\ref{#1}}

\ifpdf %
 \fi

\begin{document}
                            
\newacronym{CI}{CI}{conical intersection} %
\newacronym{GP}{GP}{geometric phase} %
\newacronym{BO}{BO}{Born-Oppenheimer} %
\newacronym{LVC}{LVC}{linear vibronic coupling} %
\newacronym{DOF}{DOF}{degrees of freedom} %
\newacronym{PES}{PES}{potential energy surface} %
\newacronym{DBOC}{DBOC}{diagonal Born--Oppenheimer correction} %
\newacronym{BMA}{BMA}{bis(methylene) adamantyl} %
\newacronym{FC}{FC}{Franck-Condon} %
\newacronym{CWE}{CWE}{cylindrical wave expansion}%
\newacronym{PB}{PB}{Particle in a box}%
\newacronym{BOA}{BOA}{Born-Oppenheimer approximation} %
\newacronym{BSC}{BSCs}{bound states in the continuum}

\sloppy
\title{
Exploring vibrational ladder climbing in vibronic coupling models: 
Toward experimental observation of a geometric phase signature of a conical intersection
}

\author{Hazem Daoud}
\affiliation{Department of Physics, University of Toronto, Toronto, Ontario, M5S 1A7, Canada}
\author{Loic Joubert-Doriol} 
\affiliation{Department of Physical and Environmental Sciences,
  University of Toronto Scarborough, Toronto, Ontario, M1C 1A4,
  Canada} %
\affiliation{Chemical Physics Theory Group, Department of Chemistry,
  University of Toronto, Toronto, Ontario M5S 3H6, Canada} %
  \author{Artur F. Izmaylov} %
  \email{artur.izmaylov@utoronto.ca}
\affiliation{Department of Physical and Environmental Sciences,
  University of Toronto Scarborough, Toronto, Ontario, M1C 1A4,
  Canada} %
\affiliation{Chemical Physics Theory Group, Department of Chemistry,
  University of Toronto, Toronto, Ontario M5S 3H6, Canada} %
\author{R. J. Dwayne Miller}
\affiliation{Departments of Physics and Chemistry, University of Toronto, Toronto, Ontario, M5S 3H6, Canada}
\affiliation{Max Planck Institute for the Structure and Dynamics of Matter, Luruper Chaussee 149,22761 Hamburg, Germany}

\date{\today}

\begin{abstract}
Conical intersections (CIs) have been widely studied using spectroscopic techniques. However, CIs have mainly
been identified by rapid internal conversion transitions that take place after the photoexcitation. 
Such identifications cannot distinguish various types of intersections as well as to separate the actual 
intersection from an avoided crossing. In this paper, we investigate how ultrafast IR laser pulses can be 
utilized to stimulate nuclear dynamics revealing geometric phase features associated with CIs. 
We consider two low-dimensional nonadiabatic models to obtain optimal two- and 
three-pulse laser sequences for stimulating nuclear dynamics necessary for the CI identification.
Our results provide insights on designing non-linear spectroscopic schemes for subsequent 
probes of the nuclear wavepackets by ultrafast electron diffraction techniques to unambiguously 
detect CIs in molecules.

\end{abstract}
\maketitle
\glsresetall

\section{Introduction}
\begin{figure}
  \includegraphics[width=0.4\textwidth]{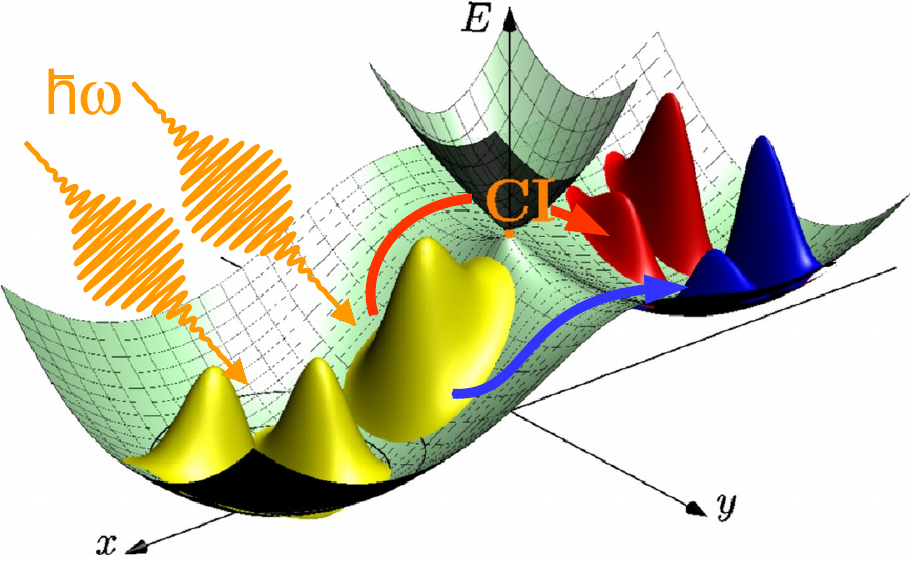}
  \caption{Destructive interference due to geometric phase in low
    energy dynamics: the initial nuclear density is in yellow and 
    the one at a later time is in red-blue.}
  \label{fig:node}
\end{figure}

Electronic potential energy surfaces (PESs) of polyatomic molecules often cross forming degenerate manifolds of nuclear configurations with the topology of conical intersections (CIs).\cite{Migani,Yarkony} Nonadiabatic dynamics associated with such crossings are of great interest to chemists and physicists for several reasons. 
CIs are the most common pathways for non-radiative transitions that drive photo-induced 
chemistry,\cite{Migani} or transfer of electronic energy and/or charge.\cite{Blancafort1,Blancafort2,Izmaylov} 
Also, CIs are associated with appearance of 
 nontrivial geometric phases (GPs) in electronic and nuclear wavefunctions of the adiabatic representation.
 GPs can profoundly affect molecular dynamics on both PESs involved in the CI.\cite{Ryabinkin:2017ch,Schon,Ryabinkin1,Baer,Ryabinkin2,Althorpe,Henshaw:2018ck,Li:2017/jcp/064106} 
 One of the most salient GP features appearing 
 in a nuclear distribution moving on a lower PES and encountering the CI is a nodal line (Fig.~\ref{fig:node}).
 This nodal line is a result of destructive interference of two parts of the nuclear distribution encircling the CI 
 from two sides and thus acquiring the opposite GPs. Note that this feature will not appear if
 two PESs are not intersecting or if the intersection is not conical (e.g. glancing intersections\cite{Yarkony:1996wx}) 
 and thus this nodal line can serve as unambiguous identification of the CI.  
 
Studying nuclear dynamics through CIs with spectroscopic methods is an active area of research,
here, usually non-linear techniques are used to probe molecular motion through CIs.\cite{Dario,Lim,Domcke,Pang} 
In pump-probe experiments, a nuclear wave packet is excited by the pump pulse from the ground PES
to the first excited PES, and the subsequent nuclear dynamics is probed by the time delayed probe pulse. 
However, due to the vanishing gap between electronic states in the vicinity of the CI, molecular dynamics near CIs cannot be optically probed. 
The laser frequency cannot be tuned to resonate with a vanishing 
energy gap between the electronic states at the CI vicinity, hence, this region is optically ``dark".
More sophisticated photon-echo pulse sequences also have not provided, so far, unambiguous 
specific signatures of the CI.\cite{Krcmar:2014ed,Krcmar:2015fh} 
Alternatively, laser light was used to modulate the electronic energy surfaces to artificially induce a conical intersection and study the associated wavepacket dynamics.\cite{Natan} 
Therefore, the existence of 
(naturally occurring) 
CIs is usually inferred by comparing experimental results for quantum yield, nonradiative relaxation dynamics and other characteristic parameters with theoretical quantum dynamics calculations.\cite{Dario} 

A promising method for the direct probe of CIs is ultrafast electron diffraction (UED).\cite{Miller1,Miller2,Sciaini} 
UED is a method that probes molecular nuclear motion at the femtosecond temporal and now achieved 0.01 angstrom
spatial scales\cite{ishikawa} by diffracting electrons.\cite{Sciaini, Daoud} Therefore, if one can prepare dynamics of a nuclear 
wavepacket toward the CI (Fig.~\ref{fig:node}), the UED technique can be used to identify the presence 
of the nodal line in the nuclear distribution and thus experimentally observe the CI. 
For example, in an experiment where the coupling mode corresponds to a bond elongation or compression, the nodal line would be observed experimentally by examining the bond length through analyzing the diffraction pattern and detecting that upon excitation the molecules never attain the equilibrium bond length of the ground state.
To initiate nuclear dynamics 
in the needed direction, we explore use of ultrafast IR lasers in the  3-10 $\mu$m range, 
which have been recently developed.\cite{elsaesser_2017, kanai_2017, sanchez_2016, kroetz} 

In order to study the feasibility 
of strong field control of molecular dynamics, we consider two simple low-dimensional two-electronic-state models:
1) one-dimensional spin-boson (1D-SB) model, and 2) two-dimensional linear vibronic coupling (2D-LVC) model.
Both of these models are formulated in the diabatic representation.\cite{Koppel:1984/acp/59} 
Although the GP itself is only present in the adiabatic representation, the effects in nuclear dynamics that are related to its existence in the adiabatic representation can be observed in other representations (e.g., diabatic) and thus are universal.\cite{Ryabinkin:2017ch}
The main advantage of working with the diabatic models is smoothness of 
all terms in the corresponding Hamiltonians.  In the case of the two models under consideration, there is also a convenience 
of using analytical perturbation theory to describe nuclear dynamics and interaction with laser light
because the involved diabatic states are represented by harmonic oscillator potentials.  

The rest of the paper is organized as follows. Section II presents formulations of the models,  
two main experimental techniques for creating ultrafast laser pulse trains, and perturbative analysis of 
nuclear dynamics stimulated by laser pulses and interstate couplings. Section III has illustrative 
dynamical simulations using exact numerical propagation.  Section IV discusses aspects of generalization 
to molecular systems with more nuclear degrees of freedom (DOF). Section V summarizes pulse 
optimization strategies to maximize the population transfer which would maximize 
the strength of the diffraction signal in a typical UED experiment. Finally, Section VI concludes 
the paper by summarizing main results and providing future outlook.

\section{Theory}
\subsection{The Model}
We begin by considering the model Hamiltonian for the two-electronic-state 
system, which will be interacting with a laser field 
\begin{equation}
\hat{H}=\hat{H}_{0}+\hat{V},
\end{equation}
where
\begin{equation}
\hat{H}_{0}= \hat{T}_{N}\bold{1}_2+
\begin{bmatrix}
V_{11} & 0\\
0 & V_{22}
\end{bmatrix},
\end{equation}

\begin{equation}
\hat{V}=\begin{bmatrix}
0 & V_{12}\\
V_{21} & 0
\end{bmatrix},
\end{equation}
$\hat{T}_{N}$ is the nuclear kinetic energy operator, $\bold{1}_{2}$ is a $2 \times 2$ unit matrix, $V_{11}$ and $V_{22}$ are diabatic potentials coupled by the $\hat V$ operator.  
We study the 1D SB and the 2D LVC models, where
\begin{equation}\hat{T}_{N}=-\frac{1}{2}\sum_{i=1}^{n}\frac{\partial^{2}}{\partial{x_{i}^{2}}}, \end{equation} 
\begin{equation} V_{11}=\sum_{i=1}^{n}\frac{\omega_{i}^{2}}{2}x_{i}^{2},\end{equation} 
\begin{equation} V_{12}=V_{21} =\sum_{i=1}^{n}o_{i}x_{i}^{i-1}, \end{equation}
\begin{equation}V_{22}=\Delta + \sum_{i=1}^{n}\frac{\omega_{i}^{2}}{2} (x_{i}-a_{i})^{2}, \end{equation}
for $n=1, 2$ for the SB and LVC models, respectively. Here $x_i$ are analogues of mass-weighted 
normal modes. We express all quantities in atomic units. 
For the 2D-LVC model, the $x_1$ coordinate separates minima of the diabatic potentials 
$V_{11}$ and $V_{22}$ and is usually referred to as the tuning coordinate, while the $x_2$ coordinate contributes to 
the coupling between the diabatic states, $V_{12}$, and is thus referred to as the coupling coordinate.  

\begin{figure}[h]
	\centering
	\includegraphics[width=0.4\textwidth]{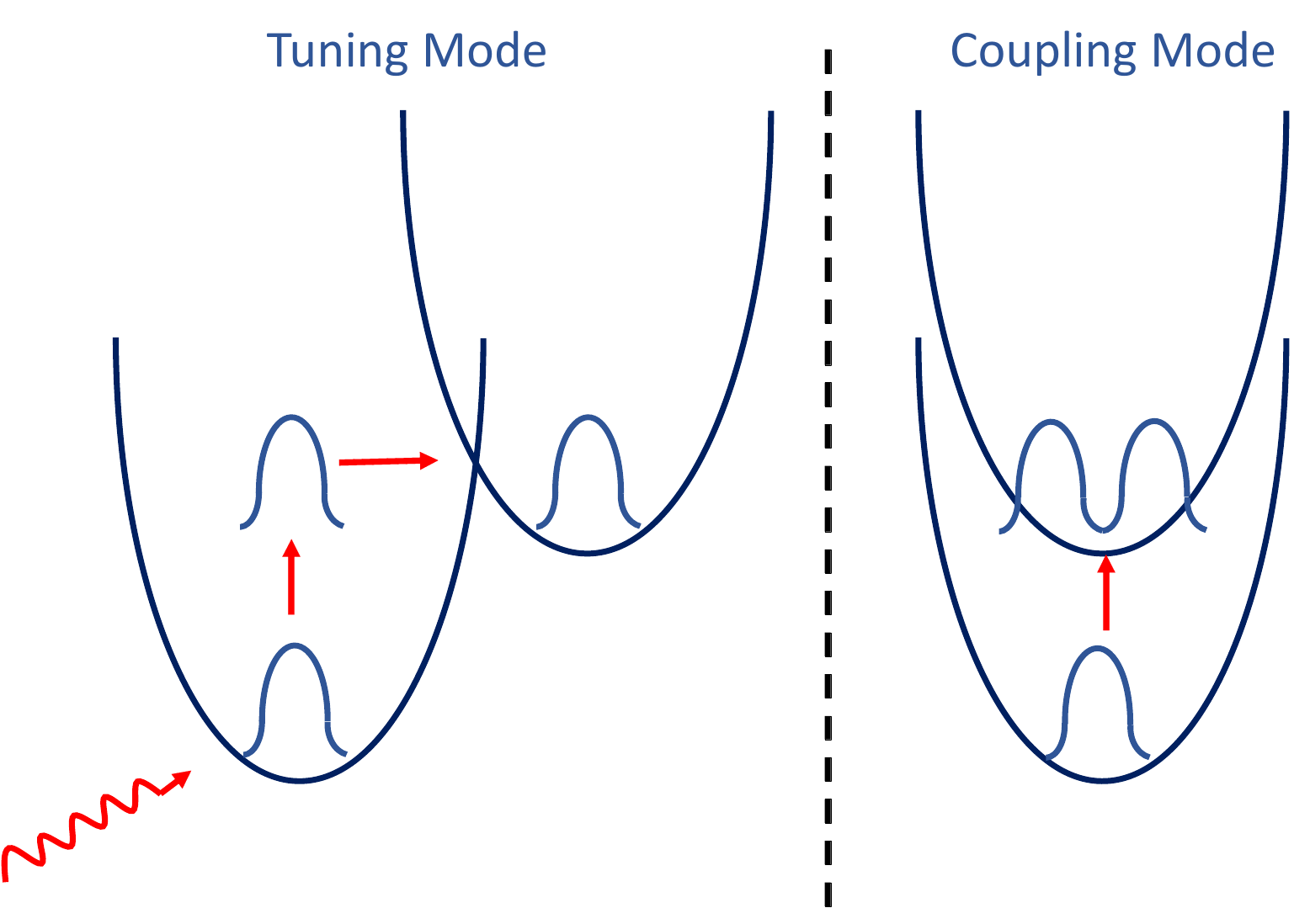}
	\caption{Schematic of the pumping process for observation of the CI nodal signature. A molecule in the lower diabatic electronic state is pumped up and crosses to the higher electronic state. In the tuning mode, there is no restriction on the shape of the nuclear probability density function (blue) of the molecule. In the coupling mode a node is formed due to the CI.}
	\label{fig:node_d}
\end{figure}

There are well defined transformations from all diabatic models to their adiabatic counterparts that involve 
diagonalizing the potential matrix $\hat V$. The eigenvalues of $\hat V$ have a dependence on the nuclear DOF and 
form PESs. The 1D-SB and 2D-LVC models give rise to avoided crossing and conical intersection in their 
adiabatic PESs, respectively. The lowest PES for both models can be made to have a double-well structure by appropriate 
selection of parameters. The nuclear dynamics that is of interest to  probe the CI requires a nuclear 
wave-packet to go from one well to another in the adiabatic representation. The same dynamics 
corresponds to wave-packet switching between two diabatic states. Further connection between dynamics 
in the two representations can be made: the nodal formation due to destructive interference induced by GP in the adiabatic representation has an origin in the linear coupling $V_{12}$ of the 2D-LVC diabatic model.\cite{Ryabinkin1}
If one prepares a nuclear wave-packet on $V_{11}$ and nodeless in the $x_2$ direction, then during 
system transfer to the $V_{22}$ state due to $V_{12}$ coupling such a wave-packet will acquire the nodal 
line in the $x_2$ direction (Fig.~\ref{fig:node_d}). 
  
\subsection{Pulsing Techniques}

To examine controlling the transfer between diabatic electronic states we examine 
two distinct schemes in regard to inter-pulse time delays and pulse phases 
that correspond to the two main experimental methods for generating 
time delayed laser pulses.\cite{Albrecht}

\textit{Collinear phase-coherent pulse (CPCP) train:}
 For CPCPs, the light-matter interaction Hamiltonian is  
 \bea
 \hat{H}_{P}(t)=\cos(\omega t+\theta) \sum_{i=1}^{N} \mathcal E_{0} e^{-\frac{(t-T_{i})^{2}}{2\tau^{2}}}\vec{\epsilon}\vec{\mu}\bold{1}_{2}
\eea 
where $N$ is the number of Gaussian laser pulses, 
$T_{i}$ is the central time of the $i^{th}$ laser pulse, 
$\mathcal E_{0}$ is the peak electric field strength, 
$\omega$ is the laser frequency, $\tau$ is the rms pulse duration, 
$\vec{\epsilon}$ is the polarization unit vector and 
$\vec{\mu}$
is the dipole moment vector. 
 $\hat{H}_{P}$ corresponds to a train of pulses with fixed phase $\theta$.  
Experimentally, different techniques exist to produce CPCP trains, the simplest of which is to split an ultrashort pulse by beam-splitters and force the different parts of the beam to travel different distances before reaching the sample. The path difference should be an integer multiple of the laser wavelength.\cite{Xu,Warren, Cheng,Tan,Mukamel,Albrecht,Keusters} \\ 

\textit{Collinear phase-delay-coupled pulse (CPDCP) train:} \\
In contrast to CPCP trains, CPDCP trains exhibit a coupling between the inter-pulse 
time delay and the carrier wave phase. Its light-matter interaction Hamiltonian is 
\bea
\hat{H}_{P}(t)=\sum_{i=1}^{N}\cos(\omega(t-T_{i})+\theta) \mathcal E_{0} e^{-\frac{(t-T_{i})^{2}}{2\tau^{2}}}\vec{\epsilon}\vec{\mu}\bold{1}_{2}.
\eea
Here, each pulse accumulates an extra phase $-\omega T_{i}$ 
so the carrier phase is a function of the inter-pulse time delay.\cite{Tan,Albrecht,Tekavec,Jonas} 

\subsection{Perturbative Analysis of Population Transfer Dynamics}

To simplify the problem, we divide the population dynamics in two processes: pumping dynamics and inter-electronic state transfer dynamics. The two processes are generally ordered in time, i.e. first, the molecule is pumped to a resonant level, and then the transfer to the other electronic state follows.

\paragraph{Pumping Dynamics:}
We study the population response to two resonant laser pulses. We start by analyzing the results of first order perturbation theory after each pulse in order to investigate the effect of the phase relationship between the pulses on the dynamics. In subsequent sections, we will use the general phase relationships derived to clarify how CPCPs and CPDCPs affect the dynamics.

We start with the entire population in the ground vibrational state $\ket{\chi_0}$ of $V_{11}$
\bea
\ket{\Psi(t=0)} = \ket{1}\ket{\chi_0},
\eea
where $\ket{1}$ is the diabatic electronic state corresponding to $V_{11}$.
The pulse analysis will be done within the first order of time-dependent perturbation theory because 
it is a dominant order and it allows for a simple illustration of trends that appear in nonperturbative 
simulations as will be discussed later. We denote the vibrational state coefficients obtained within the first order PT 
by $c_{k}^{(n)}$ where $k$ enumerates the vibrational states $\ket{\chi_k}$ and $n$ corresponds to the pulse number. The state energies of $\hat H_0$ are $E_{k}$ and the phases of the coefficients $c_{k}^{(n)} = |c_{k}^{(n)}|\exp{(i\phi_k^{(n)})}$ are $\phi_{k}^{(n)}\in(-\pi,\pi]$. 
We study the evolution of the first excited state population by finding the first order perturbative correction due to a resonant Gaussian laser pulse
\begin{equation}\dot{c}_{1}^{(1)}=\frac{1}{i} P_{10}(t)e^{i\omega_{10}t}\end{equation}
where
\begin{equation} P_{10}(t)=\left\langle \chi_1\left|\bra{1}\hat P(t)\ket{1}\right|\chi_0\right\rangle, \end{equation}
\begin{equation}\omega_{10}=E_{1}-E_{0},\end{equation}
\begin{equation}\hat P(t)=\mathcal {E}_{0}e^{-\frac{(t-T_{1})^{2}}{2\tau^{2}}}\cos(\omega_{10}t+\theta_{1})\mu\mathbf{1_2}. \end{equation}

The first order perturbative correction after the pulse has passed, i.e. for $t>T_{1}+3\tau$, and for narrow-bandwidth $(\omega>\frac{1}{\tau})$, which is the case of consideration in this study, is 
\begin{equation}c_{1}^{(1)}\cong\frac{\mu \mathcal {E}_{0}}{2i}\sqrt{\frac{\pi}{2}}\tau e^{-i\theta_{1}} \label{equ12} \end{equation}
The effect of the first pulse is to excite a fraction of the population from the ground state to the first excited state. Moreover, the phase of the coefficient of the first excited state depends on the laser pulse carrier phase $\theta_{1}$.\\

The perturbative correction due to the second pulse that follows the first pulse with a time delay $\Omega$ is
\begin{equation}c_{1}^{(2)}\cong\frac{\mu \mathcal {E}_{0}}{2i}\sqrt{\frac{\pi}{2}}\tau e^{-i\theta_{2}} \label{equ13} \end{equation}
Adding both corrections the result is
\bea
c_{1}&=&c_{1}^{(1)}+c_{1}^{(2)} \\
&=&\frac{\mu \mathcal {E}_{0}}{2i}\sqrt{\frac{\pi}{2}}\tau (e^{-i\theta_{1}}+e^{-i\theta_{2}})\\
&=&\frac{\mu \mathcal {E}_{0}}{2i}\sqrt{\frac{\pi}{2}}\tau e^{-i\theta_{2}}(1+e^{i(\theta_{2}-\theta_{1})}) \label{equ14} 
\eea

These equations show the following:\\
1.	The first pulse always excites a certain fraction of the ground state population to the first excited state.\\
2.	The second pulse can excite more population to the first excited state (absorption) and 
can cause deexcitation from the first excited state to the ground state (stimulated emission) to various degrees based on the relative carrier phase between the first and second pulse. If the pulses are in phase, i.e. the relative phase $\theta_{2}-\theta_{1}=0$, both pulses have a net effect of excitation. If $\theta_{2}-\theta_{1}=\pi$ then the second pulse has a net effect of stimulated emission.\cite{tannor}

\paragraph{Inter-electronic state dynamics:}
The population transfer dynamics across the electronic states can be studied using perturbation theory due to the relatively weak coupling strength between electronic states. As an initial condition, at $t=0$, we let the entire population be in some vibrational state in the lower diabatic electronic potential $V_{11}$ that we denote state 
$\ket{\chi_a}$, as is typically the initial condition for systems at cryogenic temperatures ($\leq$93.15 K) or, to a good approximation, even at room temperature for the systems of consideration with resonant frequencies in the mid-IR range (3-10 $\mu$m).  
According to first order perturbation theory the transfer probability to another vibrational state in the 
electronic potential $V_{22}$, denoted as $\ket{\chi_b}$, is 
\bea
P_{ba} (t) = \frac{4|H^{'}_{ba}|^2}{\omega^{2}_{ba}} \sin^{2}\left(\frac{\omega_{ba}t}{2}\right), 
\eea
where $H^{'}_{ba}=\langle \chi_b|V_{12}|\chi_a \rangle.$
Hence, considerable population transfer takes place only to vibrational states 
separated by a small energy gap (near-resonant), 
$\omega_{ba}\sim 0$. This allows for approximating the system as an effective two state system. 
In the effective two state system we denote the vibrational state in $V_{11}$ state $\ket{\chi_a}$ and denote the corresponding resonant vibrational state in $V_{22}$ state $\ket{\chi_b}$. Initially the entire population is in state $\ket{\chi_a}$. The coefficients of the vibrational states in the effective two-state system evolve according to
\bea 
c_{a}(t)&=&\cos(H^{'}_{ba}t),\\ 
c_{b}(t)&=&-i\sin(H^{'}_{ba}t).
\eea
The phases of the coefficients undergo periodical changes with the period $2\pi/|H^{'}_{ba}|$, so
\begin{eqnarray}\label{equ17}
\phi_{a}=\left\{
\begin{array}{@{}ll@{}} 
0, & t\in \left((2n+\frac{3}{2}) \frac{\pi}{|H^{'}_{ba}|},(2n+\frac{5}{2}) \frac{\pi}{|H^{'}_{ba}|}\right) \\ 
\pi, & t\in\left((2n+\frac{1}{2}) \frac{\pi}{|H^{'}_{ba}|},(2n+\frac{3}{2}) \frac{\pi}{|H^{'}_{ba}|}\right) 
\end{array}\right.
\end{eqnarray}
and
\begin{equation}\label{equ18}
\phi_{b}=\left\{
\begin{array}{@{}ll@{}} 
-\frac{\pi}{2}, & t\in \left(2n\frac{\pi}{|H^{'}_{ba}|}, (2n+1) \frac{\pi}{|H^{'}_{ba}|}\right) \\ 
\frac{\pi}{2}, & t\in \left((2n+1)\frac{\pi}{|H^{'}_{ba}|}, (2n+2) \frac{\pi}{|H^{'}_{ba}|}\right)
\end{array}\right.
\end{equation} 
for $n=0, 1, 2, ...$ Similarly,
if state $\ket{\chi_a}$ and state $\ket{\chi_b}$ are not exactly resonant, 
with $E_{b}>E_{a}$, the coefficients would evolve according to
\bea
c_{a}(t)&=&e^{-it\omega_{ba}/2}[\cos(\beta t)+i \frac{{\omega_{ba}}}{2\beta}\sin(\beta t)], \label{equ19} \\
c_{b}(t)&=&-i\frac{H^{'}_{ba}}{\beta} e^{it\omega_{ba}/2}\sin(\beta t) \label{equ20},
\eea
where
\begin{equation}\beta=\sqrt{\frac{\omega_{ba}^{2}}{4}+|H^{'}_{ba}|^{2}} \label{equ21}.\end{equation}
For near-resonant states the characteristic timescale for phase variation is of order $2\pi/\beta$.

\textit{III. Full population transfer dynamics}\\
The entire population transfer dynamics consists of the pumping dynamics and the inter-electronic state dynamics. This division of the dynamics into two distinct phases facilitates studying SB and 2D LVC systems that are subject to resonant Gaussian laser pulses. 
In the first phase, according to \eq{equ12}, the first pulse would always pump a certain fraction of the population to the first excited vibrational state and according to \eq{equ14} subsequent CPCP pulses are always phased for further excitation unless the vibrational coefficients are phase-shifted. 
By pumping the population to a resonant state the phases of the resonant states' coefficients change with a period $2\pi/|H^{'}_{ba}|$ according to \eqs{equ17} and (\ref{equ18}). This introduces a timescale into the problem. Since the coefficients of the resonant states are changing with time, the pulses will not always be phased for further excitation so they need to be timed properly to enhance the excitation to the resonant state and in turn the transfer to the higher electronic state. For near-resonant states the timescale is of order $2\pi/\beta$, as indicated in \eq{equ19}. 
The characteristic timescales depend on the resonant states, and hence, while in the systems 
we study the energy gap $\Delta$ is comparable in magnitude to the resonant frequency $\omega$, 
the results derived are applicable for systems where $\Delta\gg\omega$. 
The main difference between the SB and the 2D LVC model is the coupling. In the 2D LVC model the coupling is linear, which imposes the restriction that the ground state in the coupling mode only couples to the first excited state. 
We study both systems using two and three pulses to stay within the limits of experimentally viable parameters. When timed correctly, more pulses would generally correspond to higher population transfer since each pulse would excite a higher fraction of the population to the resonant states, which would in turn result in higher transfer to the higher electronic state.
We also study both systems by varying the coupling strength and the energy gap $\Delta$ and establish that the relevant timescales for the problem are indeed as indicated by \eqs{equ17} - (\ref{equ21}). We carry out this analysis for CPCPs and CPDCPs and present the results in the upcoming section.

\begin{table*}
\caption{Parameters of the 1D SB and 2D LVC Hamiltonians (atomic units).} \label{tab:PARAM}
  \centering
  \begin{ruledtabular}
\begin{tabular} {ccccccccc}
 
 Model & $\omega_{1}$ & $\omega_{2}$ & $a_{1}$ & $a_{2}$ & $o_{1}$ & $o_{2}$ & $\Delta$ \\
 \hline
1D SB & 3.6500 $\times 10^{-3}$ & - & 48.500 & - & 4.9460 $\times 10^{-4}$  & - & 3.6500 $\times 10^{-3}$  \\
\hline
2D LVC & 7.7430 $\times 10^{-3}$& 6.6800 $\times 10^{-3}$ & 31.050  & 0 & 0 & 8.0920 $\times 10^{-5}$  &  1.6586 $\times 10^{-2}$  \\

\end{tabular}
  \end{ruledtabular}
\end{table*}

\section{Results and Discussion}

\subsection{Model Parameters}
For our numerical simulations, we choose the parameters outlined in Table 1. 
Most of these parameters were taken from those of diabatic Hamiltonians for pyrazine (1D-SB) and bis(methylene) adamantane radical cation (2D-LVC).\cite{Gherib}
They are suitable for our simulations because a molecule in the ground vibrational state of the lower diabatic electronic state has negligible probability of transition to the higher electronic state due to the weak coupling and off-resonance level positioning. Thus, the SB and 2D LVC models with these parameters provide an ideal opportunity to test the effectiveness of laser pulses in inducing a population transfer to the higher electronic state. The electric field parameters are $\tau=4134$ (100 fs), $\mathcal E_{0}=9.5\times10^{-3}$ which corresponds to a laser intensity of $~10^{16} \frac{W}{m^{2}}$. Realistic values of $1.062\times10^{-3}$ in the SB model and $1.502\times10^{-3}$ in the 2D LVC model for the dipole derivative 
$\partial\mu/\partial x$, \cite{Dybal} which is the leading term in the transition between vibrational states in a simple harmonic oscillator potential, are used. We simulate the exact dynamics using the $4^{th}$ order ode45 integrator implemented in the MATLAB program. 
Since we are interested in maximizing the population transfer from the lower to the higher diabatic electronic state we scan the maximum population transfer as a function of the inter-pulse time delay. The maximum population transfer for any given inter-pulse time delay is the highest population in the higher electronic state at any particular time over the course of the pulsing process. 

\subsection{1D SB Model} 

\begin{figure}[h]
	\centering
		\includegraphics[height = 7 cm]{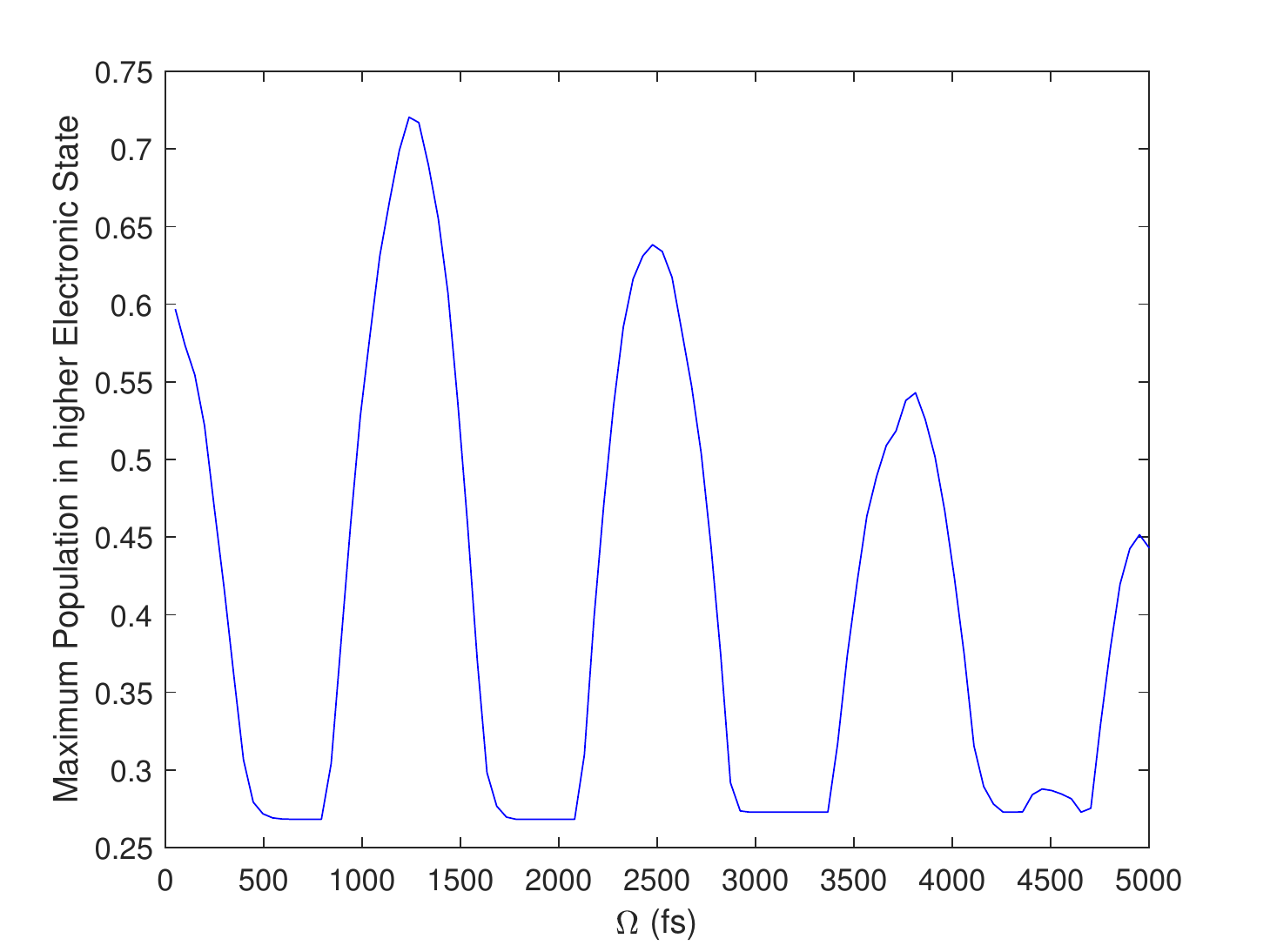}
	\caption{Maximum population transfer as a function of inter-pulse time delay $\Omega$ in the spin-boson model for 2 CPCPs.}
	\label{fig2}
\end{figure}

\begin{figure}[h]
	\centering
		\includegraphics[height = 7 cm]{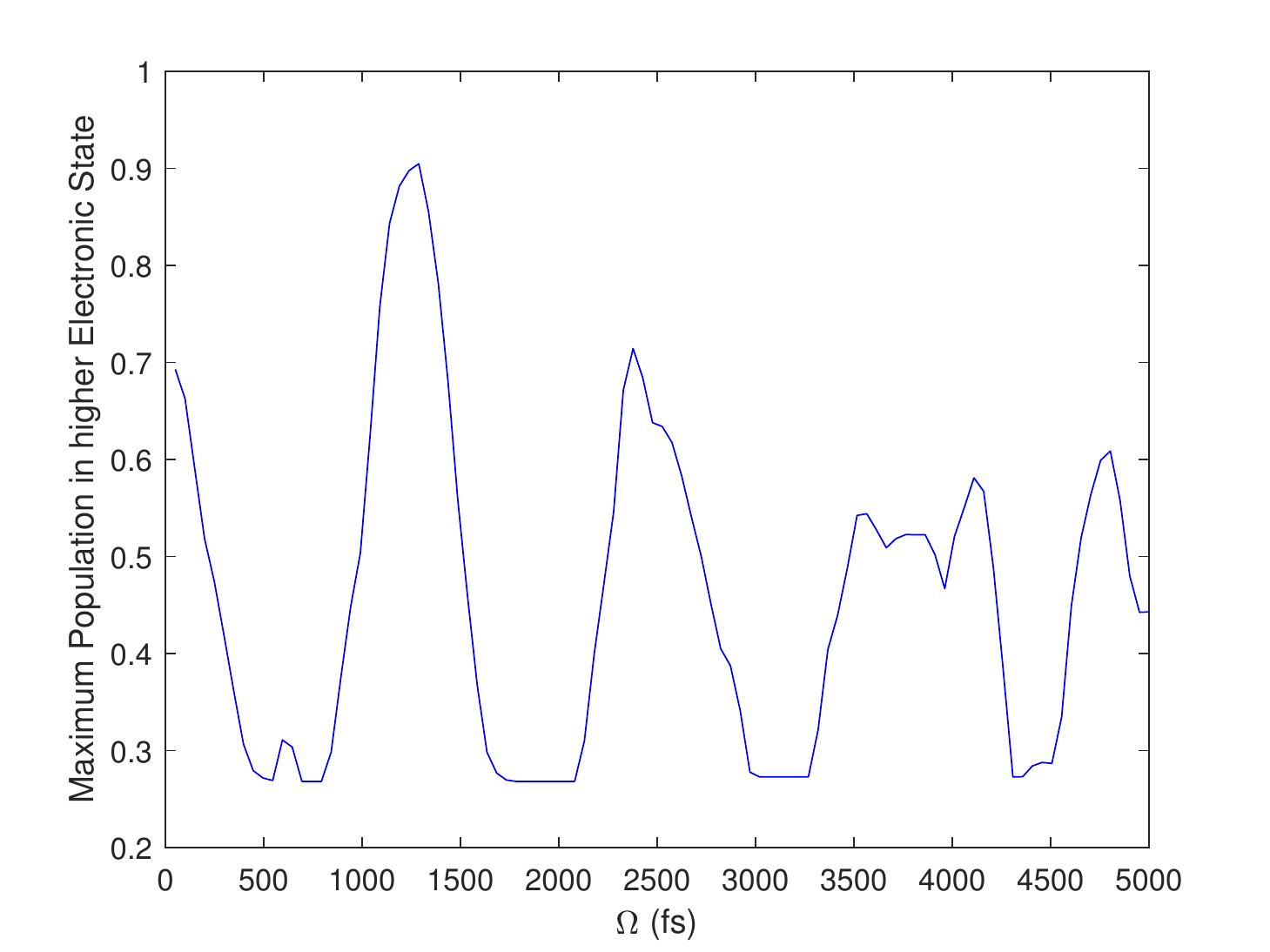}
	\caption{Maximum population transfer as a function of inter-pulse time delay $\Omega$ in the spin-boson model for 3 CPCPs}
	\label{fig3}
\end{figure}

\paragraph{CPCP train:}
Figures \ref{fig2} and \ref{fig3} show the maximum population in the higher electronic state during 
a time window $[T_{1}-4{\tau},T_{N}+20{\tau}]$ $(\tau=100$ fs$)$ as a function of the time delay between the 
pulses $\Omega$ for $N=2$ and $N=3$, respectively. The population transfer is enhanced for inter-pulse 
time delays of $1.27$ ps, $2.5$ ps and $3.8$ ps  (see Supplementary Information, Section B). 
For the deviation of $\sim 300$ fs from the optimal time delays the 
population transfer is significantly suppressed.

The maximum population transfer depends on the delay between pulses in somewhat periodic manner. 
Since the pulse train is phase-coherent, the pulse timing sensitivity is due to the periodic changes 
of the phases of the resonant states, which are of order $2\pi/|H^{'}_{ba}| \approx 1.27$ ps. 
At optimal timings, which are multiples of the characteristic period, all pulses are phased for excitation 
meaning that each pulse enhances the population excitation to the resonant vibrational state. 
For non-optimal times, the first pulse causes excitation and subsequent pulses have a net effect of 
less than optimal excitation, or de-excitation, to various degrees depending on the timing. As the pulse 
timing moves away from the optimal timing, the degree of excitation, and hence population transfer, decreases gradually which explains the amplitudes present in the graphs. The flat regions between the peaks correspond to pulse timings where the maximum population transfer takes place after the first pulse and subsequent pulses have a net effect of de-excitation and hence decrease the population transfer. 
The peaks get lower for higher time delays between the pulses as the system is not strictly a two state system in the resonant region. There exist weak couplings between all vibrational states across the diabatic electronic states. These affect the dynamics over longer time durations. 
More pulses result in stronger excitation of the population to the resonant states and hence a higher transfer to the $V_{22}$ diabatic state. 
Therefore, when timed appropriately, three pulses result in a higher population transfer than two pulses.

\begin{figure}[h]
	\centering
		\includegraphics[height = 7 cm]{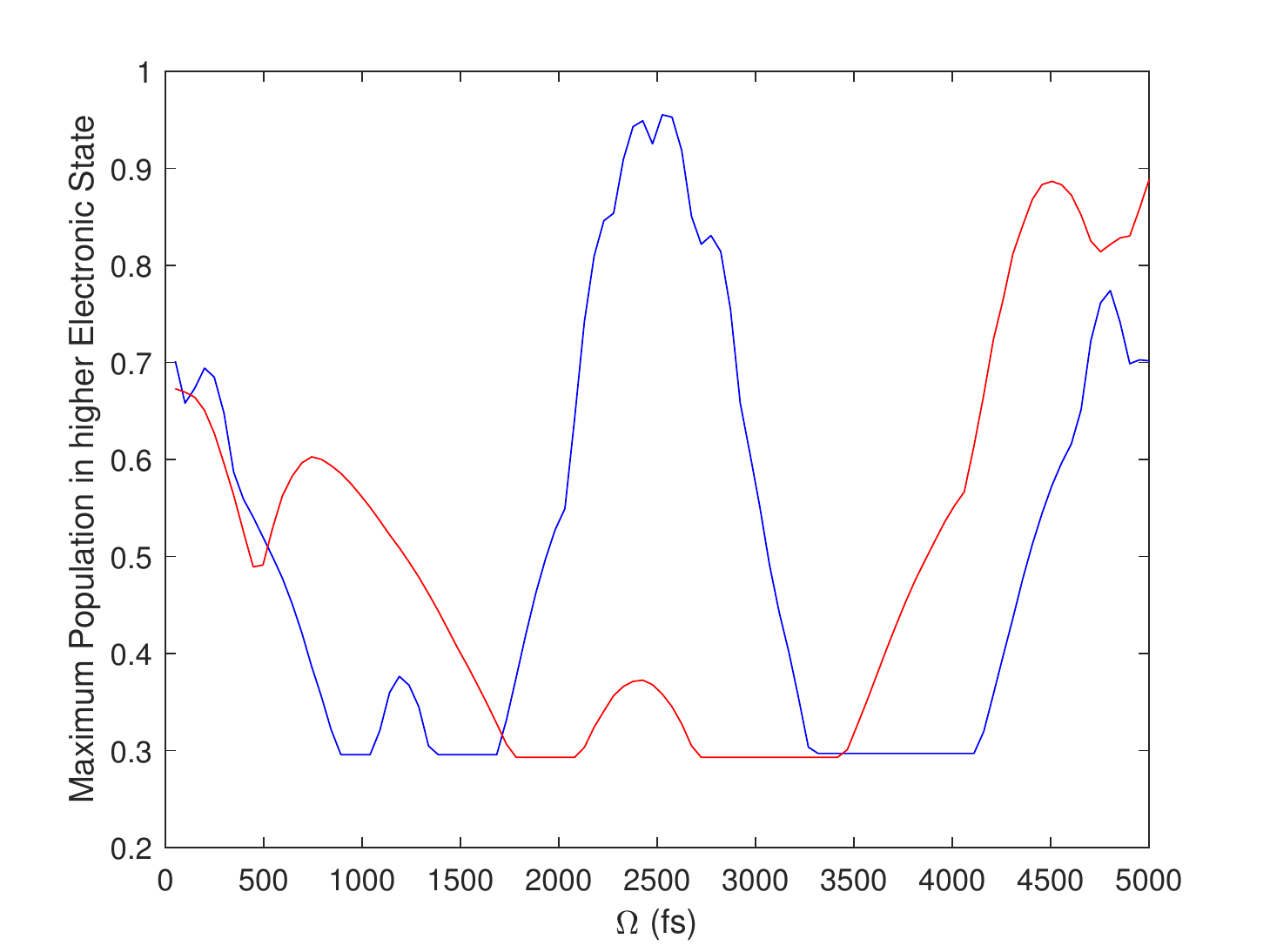}
	\caption{Maximum population transfer as a function of inter-pulse time delay $\Omega$ between pulses in the spin-boson model for 3 CPCPs for $o_{A}=0.5o_{1}$ (blue) and $o_{B}=0.25o_{1}$ (red). }
	\label{fig:cs_ch}
\end{figure}

\begin{figure}[h]
	\centering
		\includegraphics[height = 7 cm]{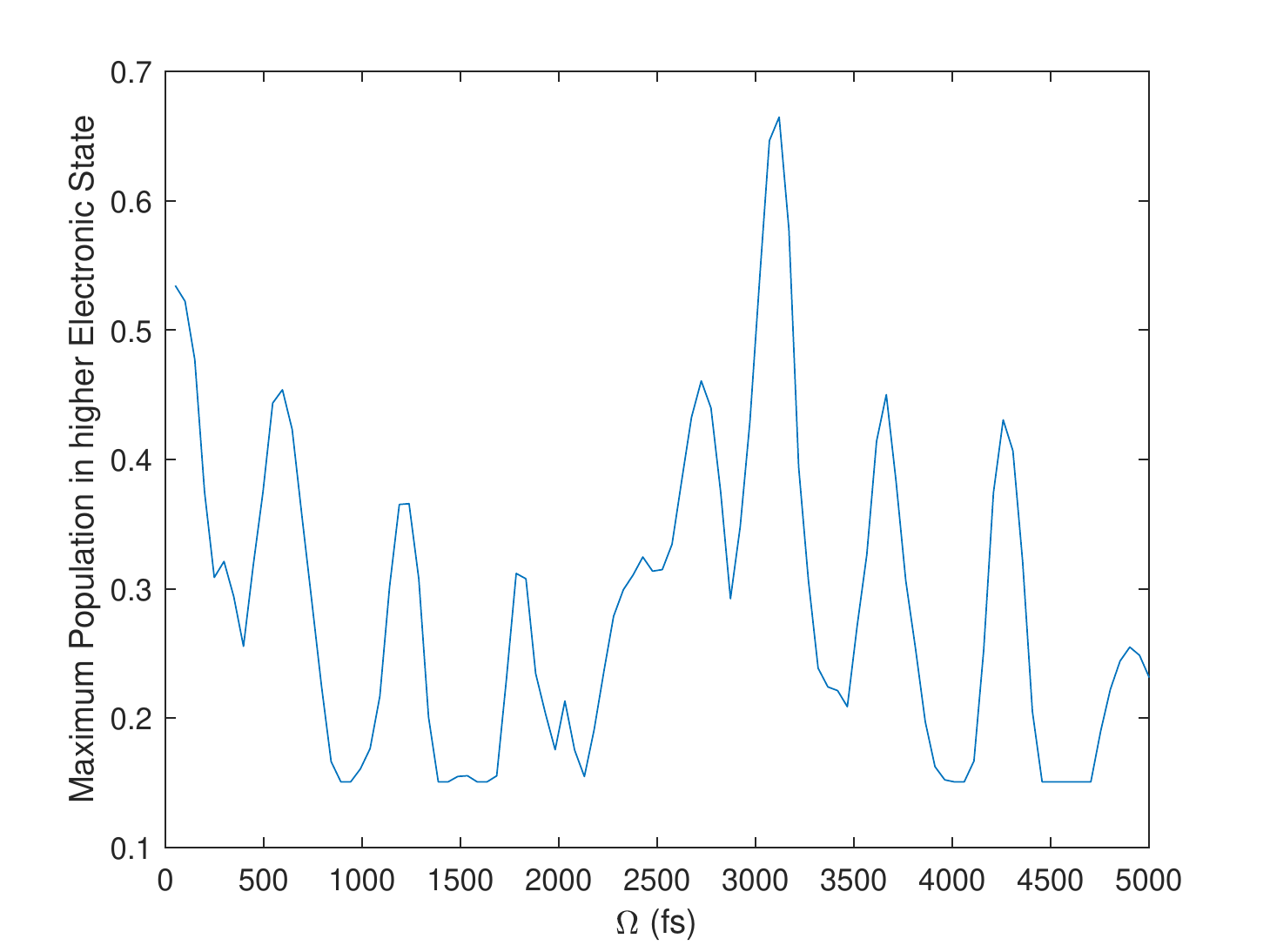}
	\caption{Maximum population transfer as a function of inter-pulse time delay $\Omega$ in the spin-boson model for 3 CPCPs for $\Delta=0.95\omega$.}
	\label{fig:en_ch}
\end{figure}

To provide further insight on the role of the coupling strength and energy gap
 between the electronic states, we did simulations with different values of these parameters 
 (Figs.~\ref{fig:cs_ch} and \ref{fig:en_ch}). 
For different coupling strengths (Fig.~\ref{fig:cs_ch}), since the period over which the phases change is inversely proportional to the coupling strength, the time delay for the first peak is doubled and quadrupled when the coupling constant is halved and quartered, respectively. 
Figure \ref{fig:en_ch} shows the maximum population transfer as a function of time delay for $\Delta=0.95\omega$. Oscillations 
over a range of $600-700$ fs 
are observed, which is consistent with calculations of a two-level system with time-independent perturbation predicting a characteristic timescale of order $2\pi/\beta$ for phase changes. 

\begin{figure}[h]
	\centering
		\includegraphics[height = 7 cm]{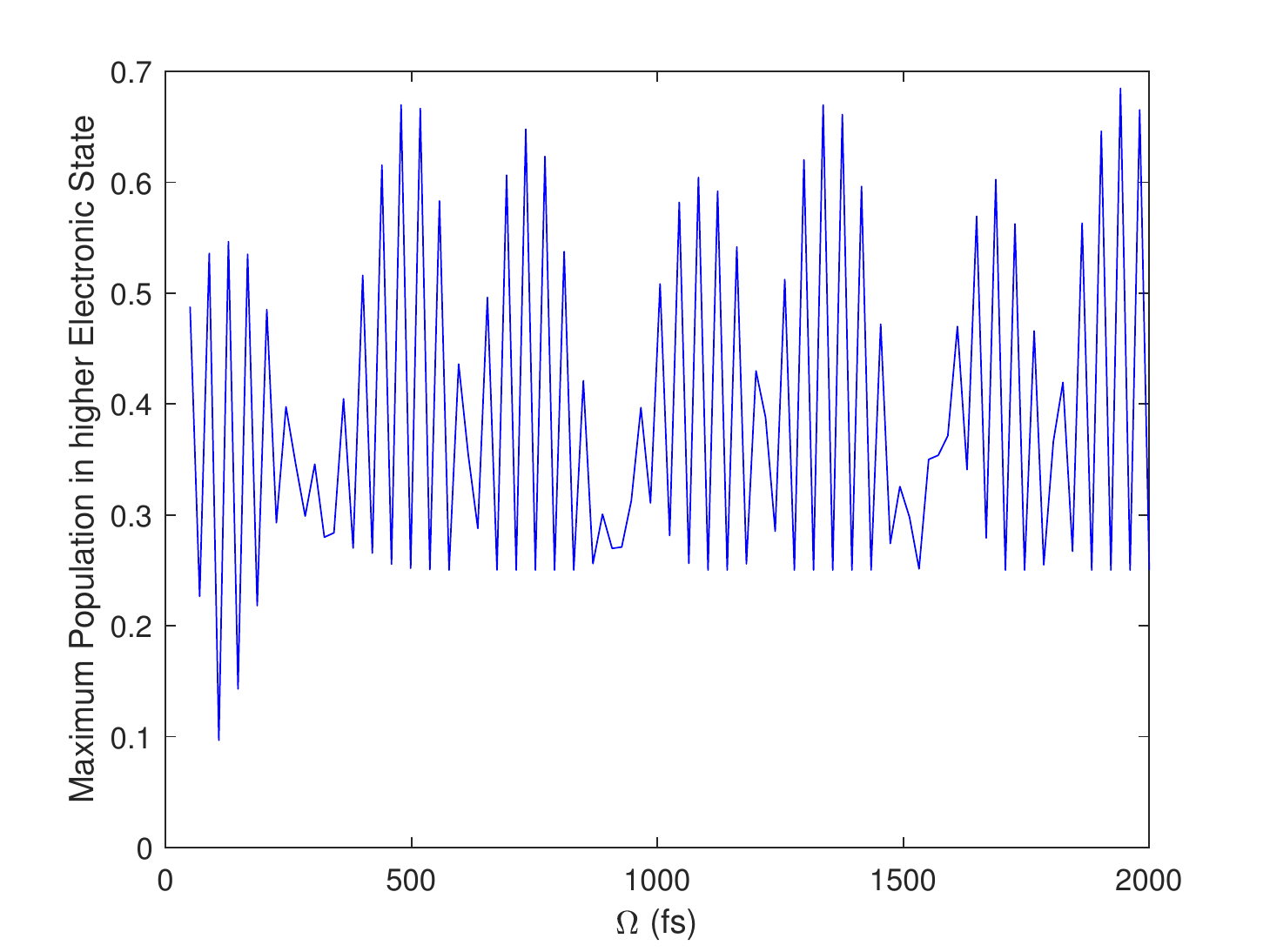}
	\caption{Maximum population transfer as a function of inter-pulse time delay $\Omega$ in the spin-boson model for 2 CPDCPs.}
	\label{fig7}
\end{figure}

\paragraph{CPDCP train:}
Figure \ref{fig7} shows the maximum population transfer as a function of inter-pulse time delay for the same SB model investigated above for two pulses. 

The sensitivity of the population transfer to the timing of the CPDCPs is much higher than for CPCPs. 
In addition to the phase changes of the coefficients of the vibrational states according to \eqs{equ17} and (\ref{equ18}), there is the added periodic change $-\omega (T_{i+1}-T_{i})$ in the phases of consecutive laser pulses over the timescale $2\pi/\omega$. This is manifested in the rapid oscillations of order $2\pi/\omega$ that are modulated by the slower oscillations of the resonant states. To obtain further insight, we varied the same parameters as for the CPCP case and examined the system dynamics. The same relationships as for CPCPs were found and details can be found in Supplementary Information, Section A.

\subsection{2D LVC Model}

We make the assumption that the laser polarization is aligned with the direction of the dipole derivative of the tuning mode and that the contribution of other modes to the dipole derivative in this direction are negligible.

\begin{figure}[h]
	\centering
		\includegraphics[height = 7 cm]{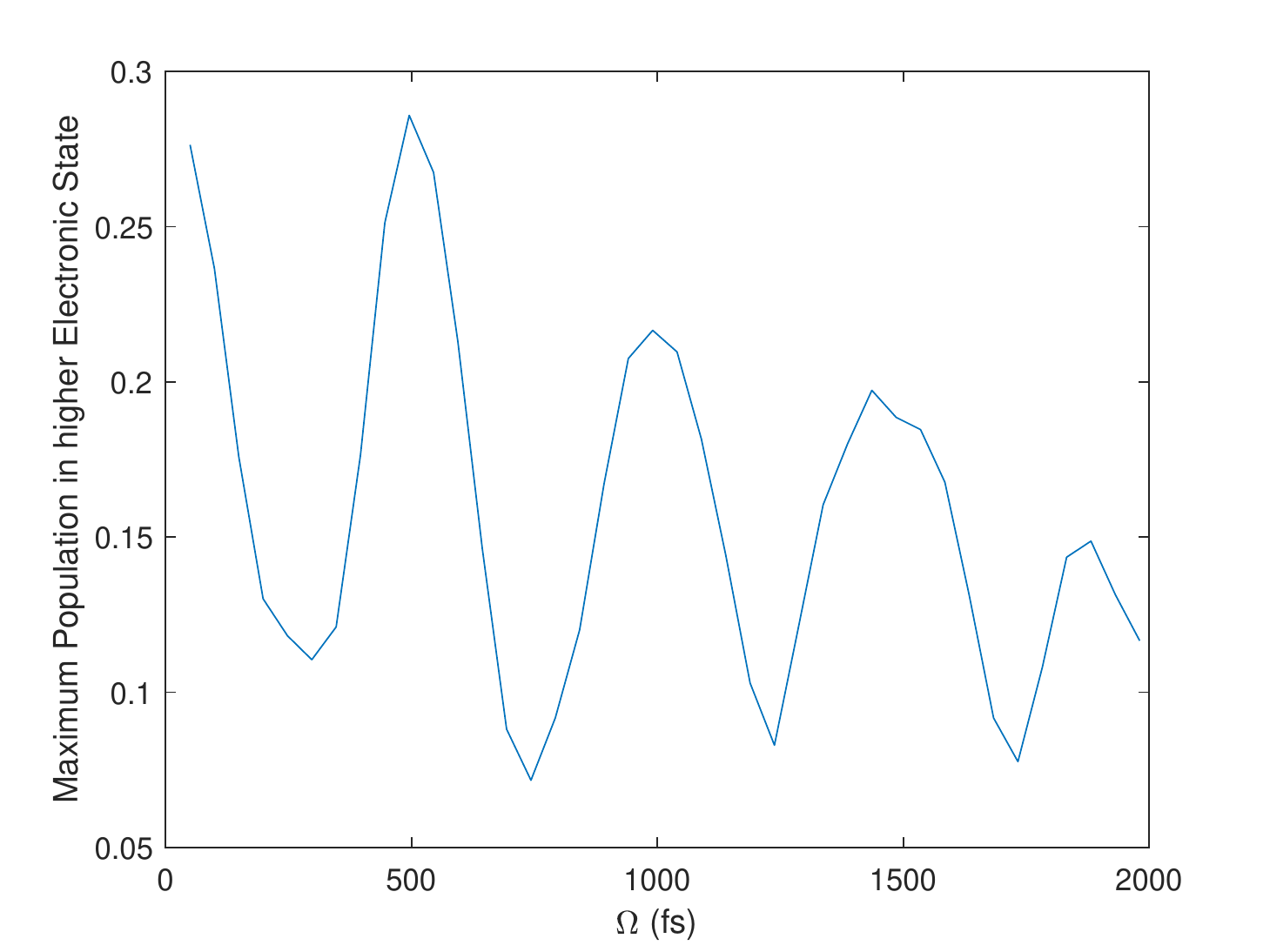}
	\caption{Maximum population transfer as a function of inter-pulse time delay $\Omega$ in the 2D LVC model for 3 CPCPs.}
	\label{fig20}
\end{figure}

\paragraph{CPCP train:} Figure \ref{fig20} shows the maximum population in the 
excited electronic state during a time window $[T_1-4\tau,T_N+20\tau]$  as a function of the time delay $\Omega$ between the pulses for $N=3$. The population transfer is enhanced for inter-pulse time delays of roughly $500$ fs, $1$ ps and $1.5$ ps (see Supplementary Information, Section B). 
For $\sim 250$ fs off the optimal time delays the population transfer is suppressed.

The same underlying physics applies to the 2D LVC model as the general resonance considerations are not affected by the coupling being linear nor by the dimensionality of the system. As in the SB model, the timing of laser pulses can either enhance or inhibit the population transfer from the lower to the higher diabatic electronic state. The timescale for phase changes of the near-resonant vibrational states is of order $2\pi/\beta$. 

\begin{figure}[h]
	\centering
		\includegraphics[height = 7 cm]{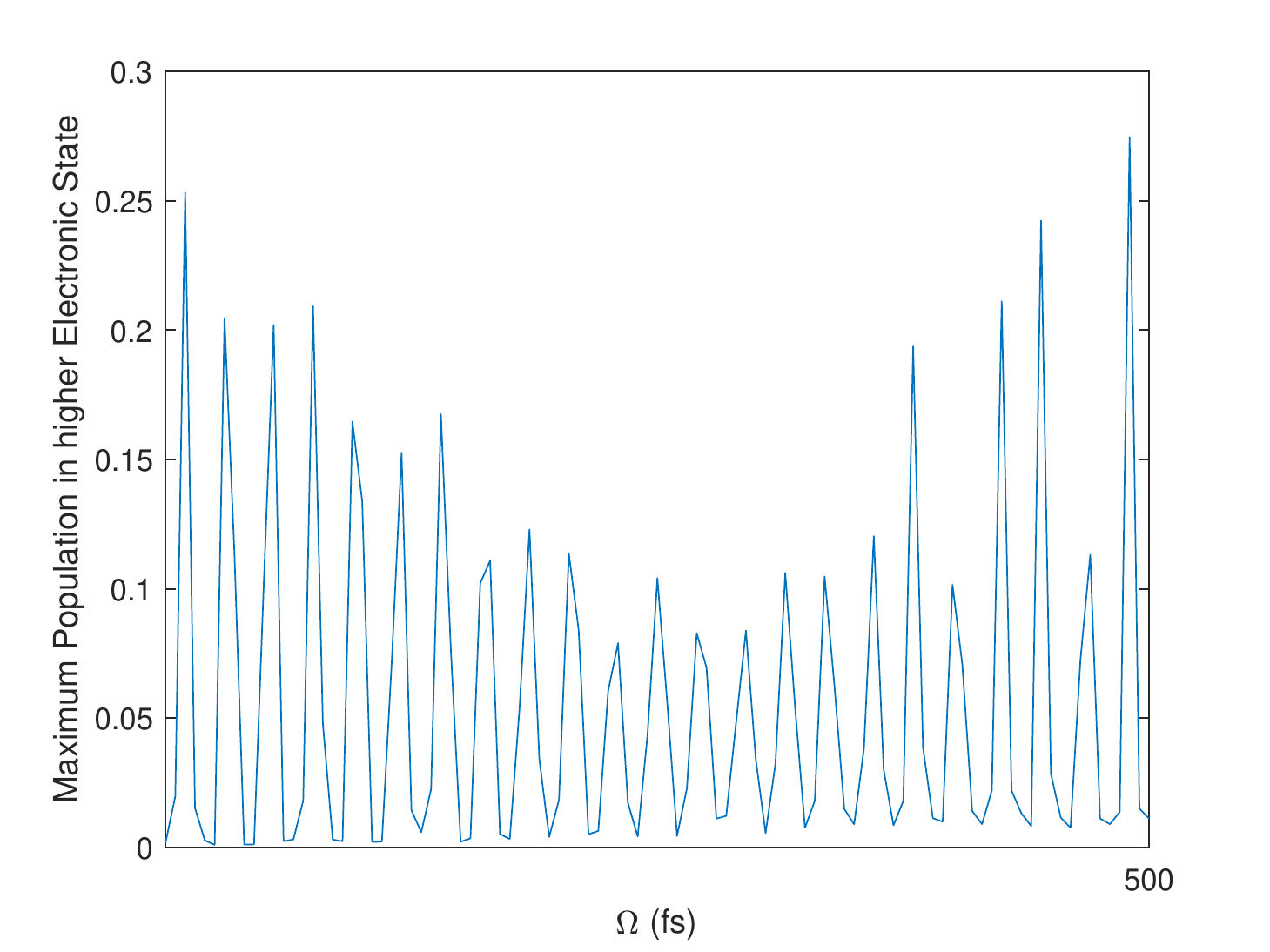}
	\caption{Maximum population transfer as a function of inter-pulse time delay $\Omega$ in the 2D LVC model for 3 CPDCPs.}
	\label{fig22}
\end{figure}

\paragraph{CPDCP train:} Figure \ref{fig22} shows the maximum population transfer as a function of inter-pulse time delay for the same model subject to three CPDCPs.

Clearly, in addition to the relatively slow phase changes of the coefficients of the resonant states, the rapid phase shifts between the pulses dominate causing the sensitivity of population transfer to inter-pulse time delay to be of order $2\pi/\omega$.

\begin{figure}[h]
	\hspace*{-1cm}
		\includegraphics[height = 7 cm, width = 10 cm]{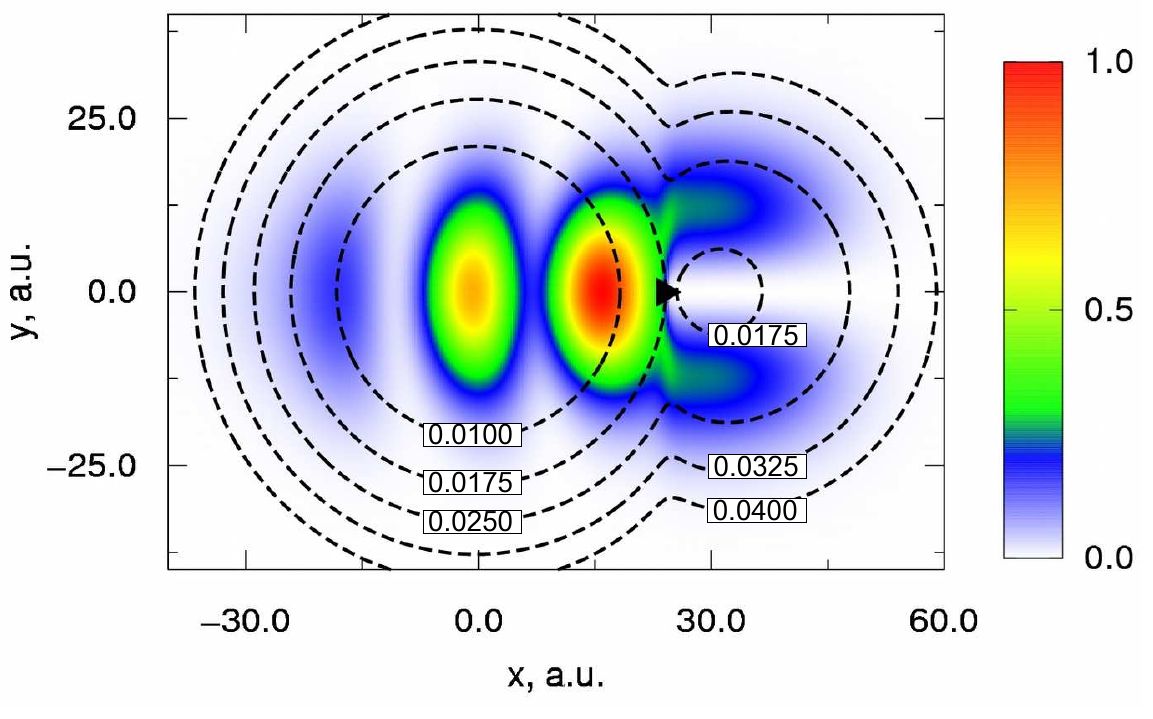}
	\caption{Color-map of the adiabatic ground state nuclear probability density at the time of maximum population transfer for three CPCPs. The dashed lines are isoenergetic contours of the electronic ground adiabatic state. The black triangle indicates the location of the CI. A nodal line in the nuclear probability density is located at $y= 0$ and $x > 24$ a.u., it clearly indicates the CI presence.}
	\label{fig_contour}
\end{figure}

Figure \ref{fig_contour} shows a contour plot of the adiabatic ground state nuclear probability density at the time of maximum population transfer for three CPCPs. A nodal line can be observed at $y=0$, a clear signature of the existence of a CI.

\section{N-Dimensional Models and Experimental Considerations}
The underlying physics discussed in this paper applies to higher dimensional LVC models as well. For CPCPs the main factors which ultimately set the timescale for the optimal inter-pulse time delays are the coupling strength and the energy difference (between the diabatic near-resonant vibrational states). 

For CPDCPs the timescale is dominated by the relative phase changes between the pulses which is of order $2\pi/\omega$. 

To extend the results derived in this paper to polyatomic molecules with multiple degrees of freedom and to help guide the choice of an appropriate system for the experiment that would feature a first direct observation of a geometric phase signature of a CI, we comment on the following factors that should be taken into account:

1. Lifetime of the nuclear vibrational states:
The lifetime of the nuclear vibrational states along the vibrational ``ladder'' should be taken into account when adjusting the degree of chirp of the pulses. Generally, the longer the nuclear vibrational states' lifetimes the easier the experiment would be. 
The wavepacket dephasing time as well as intramolecular vibrational redistribution (IVR) would also put a limit on how long the excitation process could take.
This could favor working on samples under cryogenic temperatures.  Also, samples with a rather narrow energy gap ($\Delta$) between minima of 
diabatic electronic states are preferred. 

2. Density of vibrational states in resonant region:
The density of states along with the coupling strengths to the various modes and the vibrational energy levels should be studied first to determine the characteristic timescale of resonance and hence the suitability of the system.

3. Pulse bandwidth:
Generally, tuning modes are low-frequency modes and hence pulses with a narrow bandwidth are preferred to limit the excitation of any other modes as much as possible. This is already the case for the models and pulse durations we have considered in this study.

4. Pulse polarization:
The direction of the pulse polarization should be adjusted so as to maximize the excitation along the tuning mode of consideration. In addition to a narrow bandwidth this would help limit the excitation of unintended modes. In polycrystals with random orientations, it is expected that one third of the molecules would be excited along the intended tuning mode. In mono-crystal samples the polarization direction can be deduced by studying the absorption spectrum of the sample along several directions.

\section{Optimizing the Pulsing Technique}
Based on these results we propose to divide the pulsing scheme into two phases. In the first phase a molecule in the ground state of the lower diabatic electronic state is pumped to a region of higher vibrational states that are near-resonant to vibrational states in the higher electronic state. 
This would initiate the population transfer between the electronic states. 
The second phase consists of a series of pulses at appropriate time delays to enhance the excitation to those particular vibrational states to further enhance the transfer to the higher electronic state. Experimentally, the main laser parameters to be controlled are intensity, frequency, polarization, pulse duration, inter-pulse time delay as well as the relative phases between pulses. We propose optimizing these parameters in the following way: The laser intensity should be maximized but kept below the damage threshold of the sample. 
The laser frequency should be tuned to be resonant with a tuning mode frequency. 
In a multi-step vibrational ladder climbing scheme this will require chirped laser pulses.\cite{Maas} 

The chirp speed should be set after studying the lifetimes of the vibrational states along the ``ladder". The electric field polarization should be aligned in the direction that maximizes the excitation along the intended tuning mode. The pulse duration and inter-pulse time delay should be optimized based on the timescales of phase changes of the resonant state coefficients. The pulse duration should be set to take full advantage of the time span where the phase changes of the coefficients of resonant states cause excitation in response to the laser pulses.
Time delays should be set such that the molecule would only be pulsed at times at which enhancement of the excitation would occur. The relative phase between the pulses is of utmost importance. Fixing the relative phase between all pulses, i.e. using CPCPs, generally allows for a larger margin of experimental error in the timing of the pulses. For CPDCPs the window for experimental error in the timing of the pulses would typically be much narrower. 
However, if CPDCPs are timed correctly, the maximum achievable population transfer could be reached in a considerably shorter time span than for CPCPs, which could prove useful when considering the lifetimes of the various vibrational states.

\section{Conclusion}
By considering the simple vibronic models we have identified basic physical elements crucial for detecting the 
CIs in more complex systems by their GP induced features in the nuclear density.  
Vibrational states across the electronic states that are (near-)resonant exhibit phase changes
over the ultrafast sub-picosecond timescale. These changes make the population transfer from the lower diabatic electronic state to the higher electronic state in a vibrational ladder climbing pumping scheme sensitive to the timing of laser pulses. 
We analyzed the molecular response to CPCP and CPDCP trains and discussed the different timescales corresponding to each of the pulsing techniques. 
In order to maximize the population transfer from the lower diabatic electronic state to the higher electronic state, the inter-pulse time delay for CPCPs 
should be inversely proportional to the vibronic coupling between (near-)resonant vibrational 
states at different electronic states,  
and inversely proportional to the resonant laser frequency for CPDCPs.
Based on these results we divide the pumping scheme into two phases. In the first phase the molecule is pumped rapidly to the vibrational level that is (near-)resonant with vibrational states of the other electronic state. Through this state the population transfer takes place to the other electronic state. 
In the second phase, appropriately timed subsequent laser pulses induce an enhancement of the transfer. The coherent manipulation of excited state nuclear dynamics will lead to a GP shift as the wavepacket propagates through the CI.
Our hope is that this study would eventually lead to the first direct observation of the nodal signature of CIs in a UED experiment. 

\section{Acknowledgements}

A.F.I. greatly appreciates financial support by the Alfred P. Sloan Foundation and the Natural Sciences and Engineering Research Council of Canada (NSERC) through the Discovery Grants Program. R.J.D.M. thanks the Max Planck society for support of this work.

\bibliography{mybib}

\newpage

\section{Supplementary Information}

\subsection{1D SB Model: CPDCP study} 

\begin{figure}[h]
	\centering
		\includegraphics[height = 7 cm]{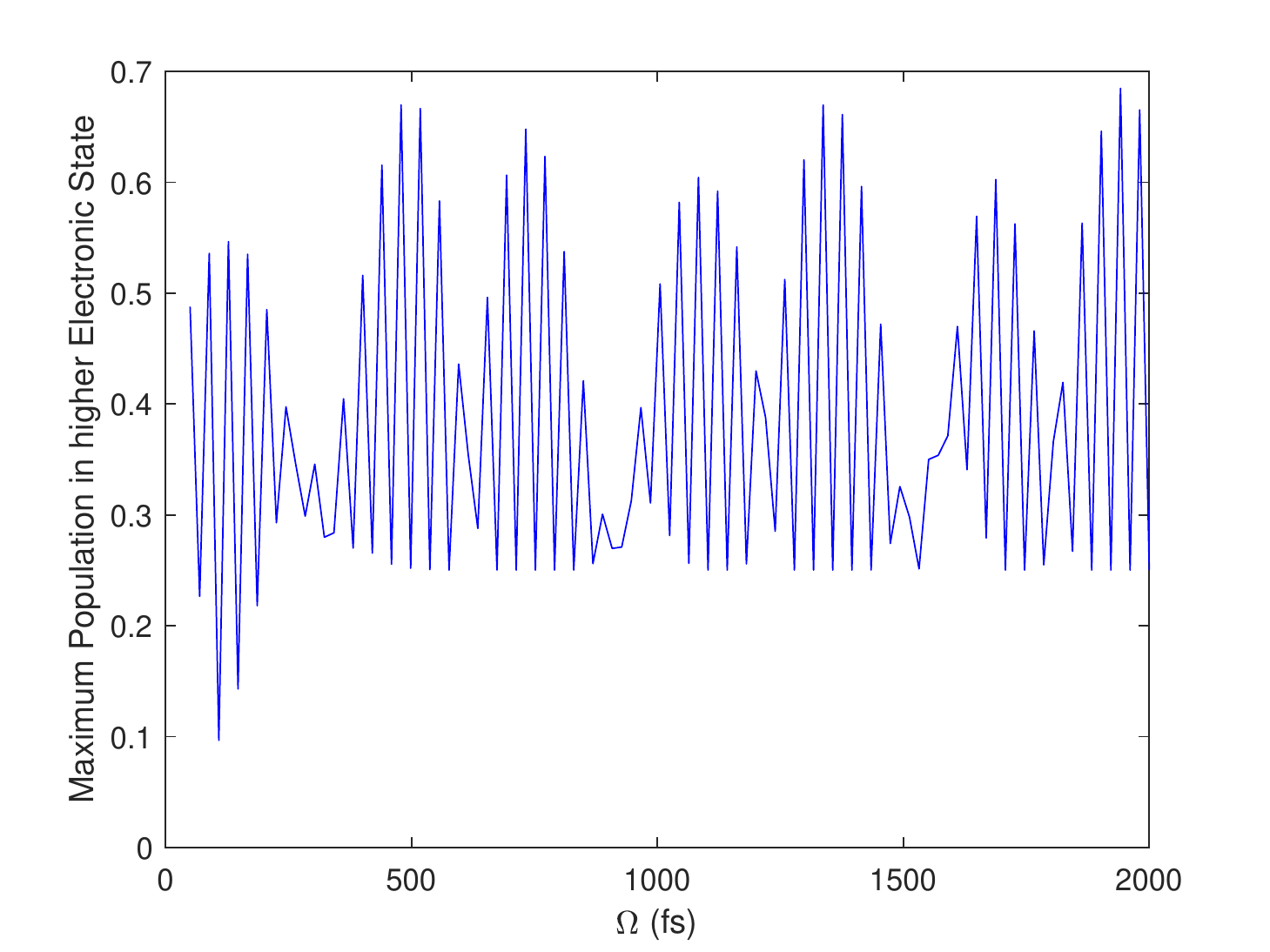}
	\caption{Maximum population transfer as a function of inter-pulse time delay $\Omega$ in the spin-boson model for 2 CPDCPs.}
	\label{fig1_supp}
\end{figure}

\begin{figure}[h]
	\centering
		\includegraphics[height = 7 cm]{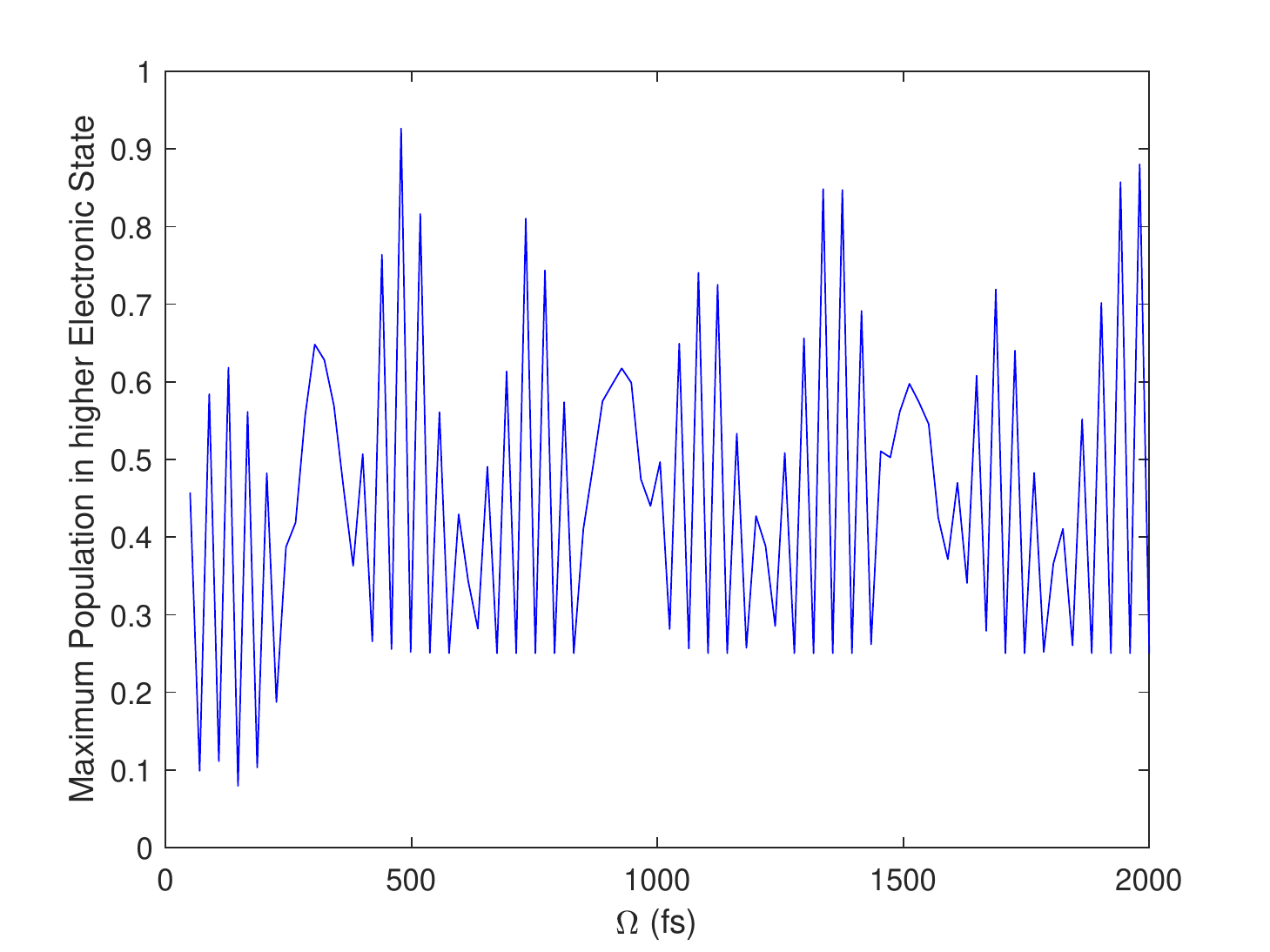}
	\caption{Maximum population transfer as a function of inter-pulse time delay $\Omega$ in the spin-boson model for 3 CPDCPs.}
	\label{fig2_supp}
\end{figure}

\begin{figure}[h]
	\centering
		\includegraphics[height = 7 cm]{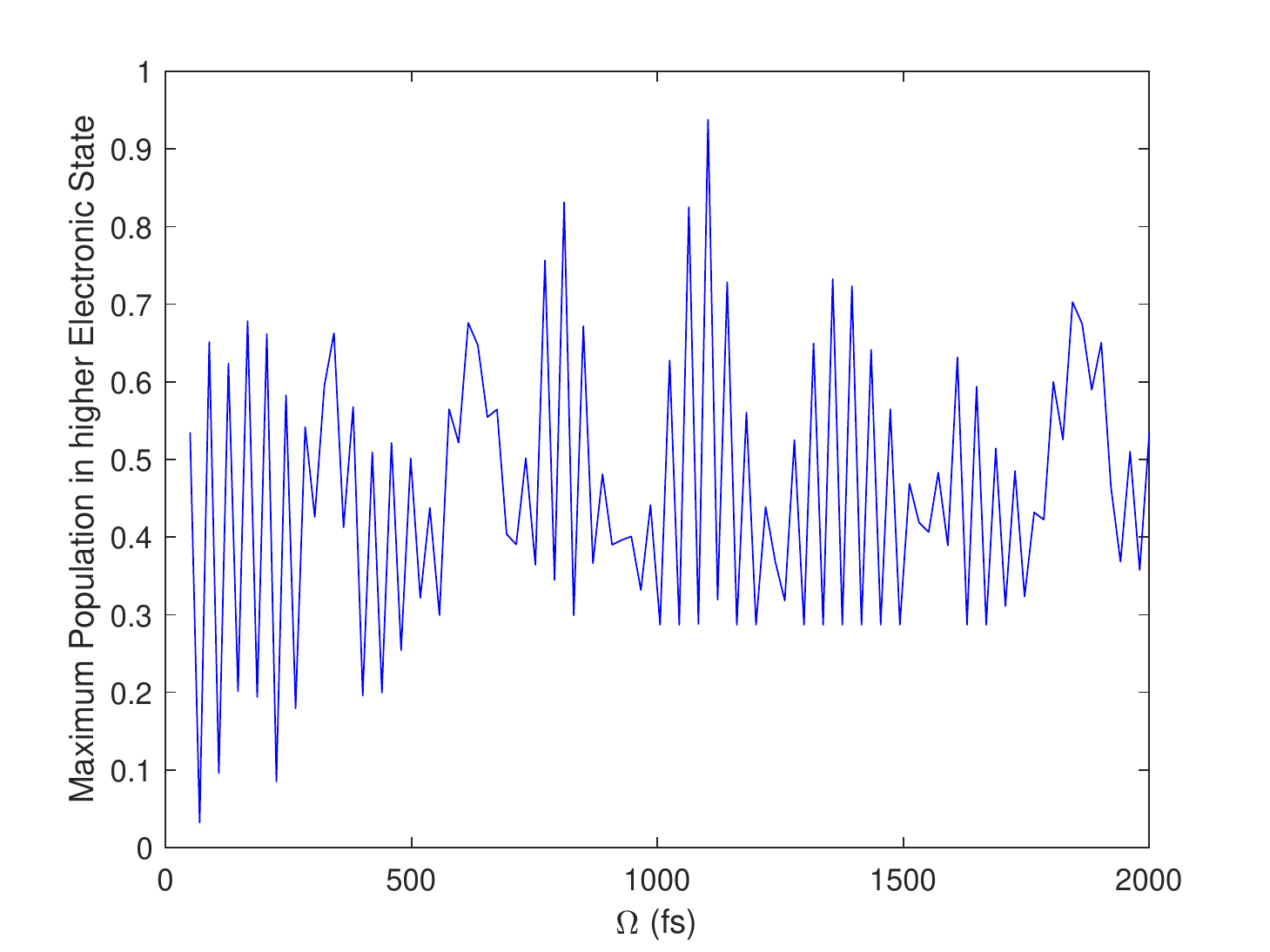}
	\caption{Maximum population transfer as a function of inter-pulse time delay $\Omega$ between pulses in the spin-boson model for 3 CPDCPs for $o_{A}=0.5o_{1}$}
	\label{fig3_supp}
\end{figure}

\begin{figure}[h]
	\centering
		\includegraphics[height = 7 cm]{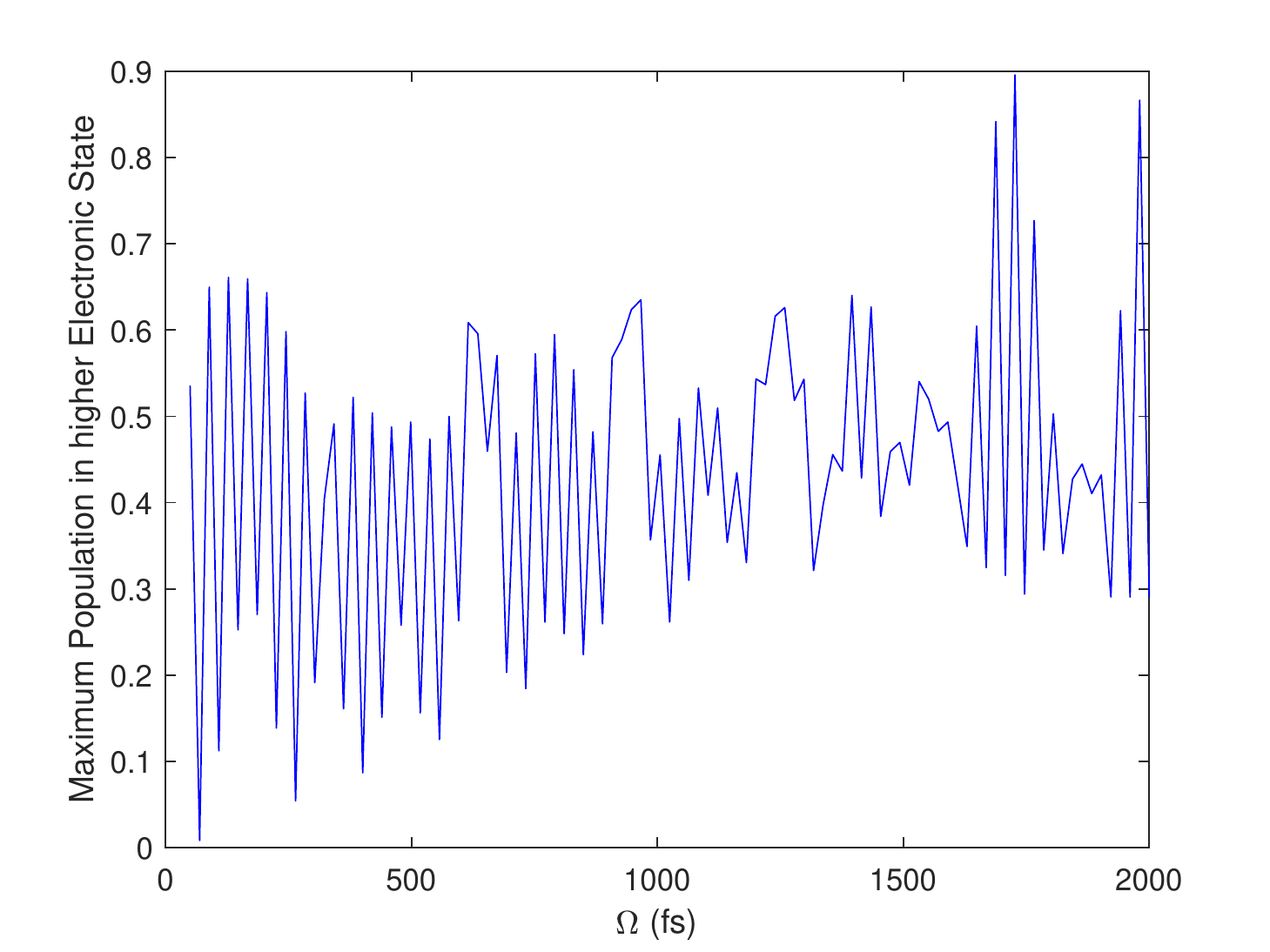}
	\caption{Maximum population transfer as a function of inter-pulse time delay $\Omega$ between pulses in the spin-boson model for 3 CPDCPs for $o_{B}=0.25o_{1}$.}
	\label{fig4_supp}
\end{figure}

\begin{figure}[h]
	\centering
		\includegraphics[height = 7 cm]{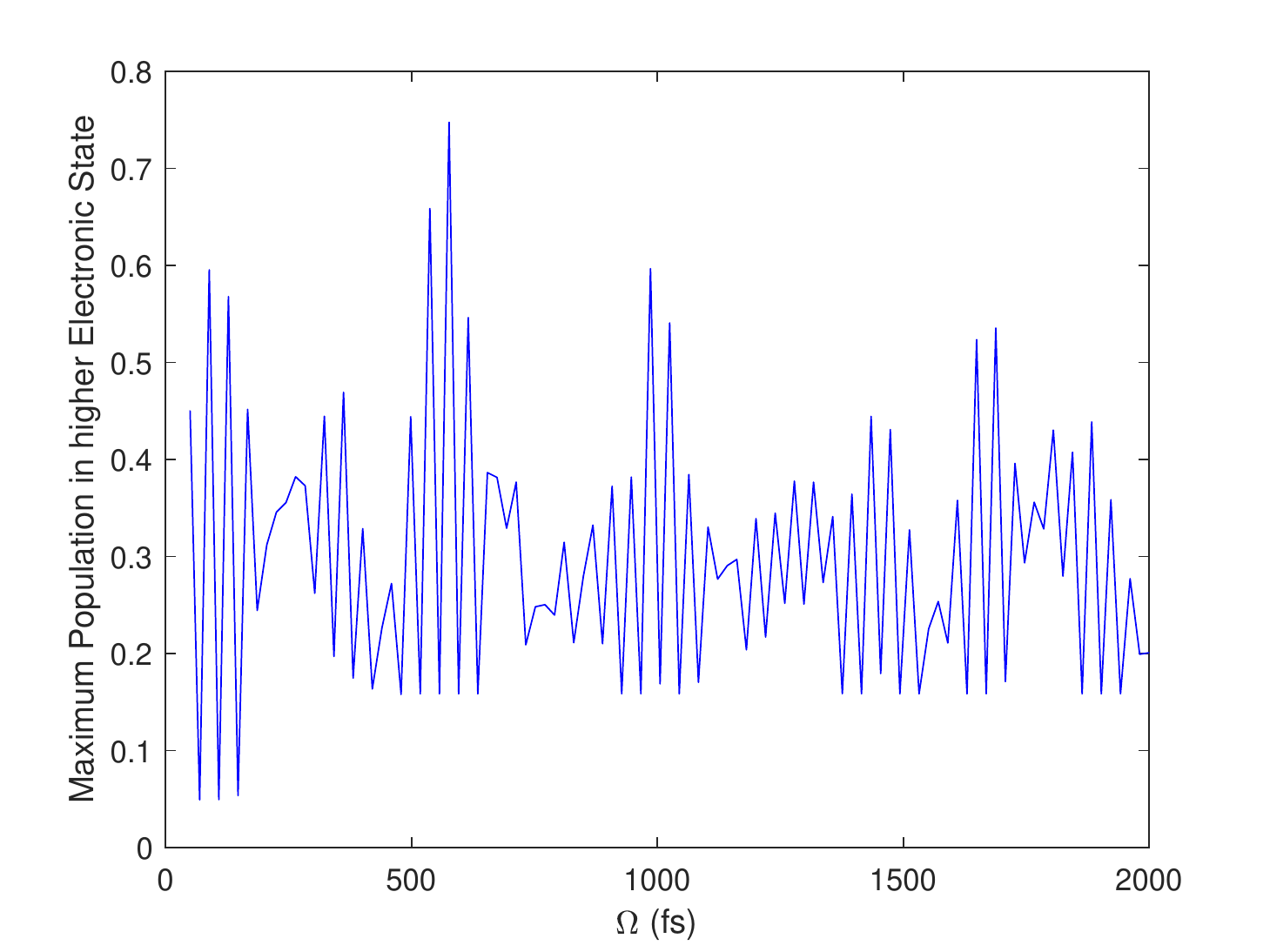}
	\caption{Maximum population transfer as a function of inter-pulse time delay $\Omega$ in the spin-boson model for 3 CPDCPs for $\Delta=0.95\omega$.}
	\label{fig5_supp}
\end{figure}

Here we summarize trends found after exploring dependence of CPDCPs results on model parameters in the 
1D SB Model. Three pulses result in a higher population transfer than two pulses (\figs{fig1_supp} and \ref{fig2_supp}). 
Weaker diabatic couplings increase the time period for maximum population transfer 
oscillations (\figs{fig3_supp} and \ref{fig4_supp}). 
The vibrational excited states in the higher electronic state being off resonance decreases the maximum achievable
 population transfer but in all cases the sensitivity of population transfer to inter-pulse time delay is dominated by
  the phase changes of the pulses rather than the resonant vibrational states (\fig{fig5_supp}). 
  These changes occur on a timescale of order $2\pi/\omega$.

\newpage

\subsection{Analytical Calculations for Characteristic Timescales}

We provide analytical calculations for the main characteristic timescales mentioned in the paper.\\
\\
I. 1D SB model (CPCPs)\\
\\
In the SB model, the first vibrational excited state in the $V_{11}$ electronic potential is resonant with ground state of the electronic potential $V_{22}$. The characteristic timescale is given by $2\pi/|H^{'}_{ba}|$.
$$ \left|H^{'}_{ba}\right|= \int_{-\infty}^{+\infty} \frac{1}{\sqrt{2}} \left(\frac{\omega}{\pi}\right)^\frac{1}{2} e^{-\frac{\omega x^{2}}{2}} 2\sqrt{\omega}x e^{-\frac{\omega (x-a_{1})^{2}}{2}} o_{1} dx$$
$$=1.198 \times 10^{-4}$$
where $\omega=0.00365$, $a_{1}=48.50$ and $o_{1}=4.946 \times 10^{-4}$ were used.
After converting to SI units, $2\pi/|H^{'}_{ba}|\approx 1268 fs$.\\
\\
Using $\omega_{ba}=0.05 \times 0.00365$ and converting to SI units we obtain $2\pi/\beta \approx 1009 fs$ 
for the non-resonant case.\\
\\
II. 1D SB model (CPDCPs)\\
\\
For CPDCPs, the characteristic timescale is $2\pi/\omega \approx 42 fs$, where $\omega = 0.00365$ was used.\\ 
\\
III. 2D LVC model (CPCPs)\\
\\
In the LVC model, the coupled vibrational levels are not resonant, thus the timescale is given by $2\pi/\beta$, 
where $\beta=\sqrt{\frac{\omega_{ba}^{2}}{4}+|H^{'}_{ba}|^{2}}$. The two vibrational states closest to each 
other are (3,0) and (0,1) where the first (second) index enumerates vibrational quanta in the tuning (coupling) mode, 
which results in $\omega_{ba}=3.759 \times 10^{-5}$. The total coupling Hamiltonian $|H^{'}_{ba}|$ is the product of the coupling Hamiltonian matrix elements for the coupling and tuning modes. We will denote them $|H^{'}_{ba}|_{c}$ and $|H^{'}_{ba}|_{t}$, respectively.\\
\\
 $$\left|H^{'}_{ba}\right|_{t}= \int_{-\infty}^{+\infty} \frac{1}{\sqrt{2}} 
\left(\frac{\omega_{1}}{\pi}\right)^\frac{1}{2}
e^{-\frac{\omega_{1} (x-a_{1})^{2}}{2}} \frac{1}{\sqrt{2^{3}3!}}$$
$$e^{-\frac{\omega_{1} x^{2}}{2}} (8(\sqrt{\omega_{1}x})^{3}-12\sqrt{\omega_{1}}x) dx=0.455429$$ 
where $\omega_{1}=0.007743$ and $a_{1}=31.05$ were used. \\
\\
$$ \left|H^{'}_{ba}\right|_{c}= \int_{-\infty}^{+\infty} \frac{1}{\sqrt{2}} \left(\frac{\omega_{2}}{\pi}\right)^\frac{1}{2} e^{-\frac{\omega_{2} x^{2}}{2}} 2\sqrt{\omega_{2}}x e^{-\frac{\omega_{2} (x)^{2}}{2}} 
o_{2} x dx$$
$$=5.72 \times 10^{-5}$$
where $\omega_{2}=0.00668$ and $o_{2}=8.092 \times 10^{-5}$ were used. \\
\\
The result is $\beta=3.194 \times 10^{-4}$. Converting to SI units, the characteristic timescale is $2\pi/\beta \approx 476 fs$.\\
\\
IV. 2D LVC model (CPDCPs)\\
\\
For CPDCPs, the characteristic timescale is $2\pi/\omega \approx 20fs$, where $\omega = 0.007743$ was used.

\end{document}